\documentclass[aps,pre,twocolumn,groupedaddress,floatfix]{revtex4}
\usepackage{amssymb,amsfonts}
\usepackage[intlimits]{amsmath}
\usepackage{graphics,epsfig}

\def\date
   {\noindent Date: \today \par
    \medskip}

\def\be{\begin{eqnarray}}
\def\ee{\end{eqnarray}}
\def\l{\left}
\def\r{\right}
\def\ni{\noindent}
\def\p{\partial}
\def\w{\omega}
\def\eps{\epsilon}

\def\Ex#1{\langle#1\rangle}

\def\EscT{{\cal T}}

\def\vth{{v_\mathrm{th}}}
\def\vre{v_\mathrm{re}}
\def\vlb{v_\mathrm{lb}}
\def\vu{v_\mathrm{u}}
\def\vs{v_\mathrm{s}}
\def\vT{v_\mathrm{T}}
\def\dT{\delta_\mathrm{T}}
\def\mu{\mathrm{u}}

\def\me{\mathrm{e}}
\def\mi{\mathrm{i}}
\def\mE{\mathrm{E}}

\def\Je{J_\me}
\def\Ji{J_\mi}

\def\Re{R_\me}
\def\Ri{R_\mi}

\def\epse{\eps_\me}
\def\epsi{\eps_\mi}
\def\betae{\beta_\me}
\def\betai{\beta_\mi}
\def\ame{a_\me}
\def\ami{a_\mi}
\def\bme{b_\me}

\def\hme{h_\me}

\def\SJ{\bar{J}}
\def\SJe{\bar{J}_\me}
\def\SJi{\bar{J}_\mi}
\def\SP{\bar{P}}
\def\Sr{\bar{r}}
\def\SRe{\bar{R}_\me}
\def\SRi{\bar{R}_\mi}

\def\Sj{\bar{\jmath}}
\def\Sje{\bar{\jmath}_\me}
\def\Sji{\bar{\jmath}_\mi}
\def\Sp{\bar{p}}

\def\MJ{\hat{J}}
\def\MJe{\hat{J}_\me}
\def\MJi{\hat{J}_\mi}
\def\MP{\hat{P}}
\def\Mr{\hat{r}}

\def\Mq{\hat{q}}
\def\Mre{\hat{r}_\me}
\def\Mri{\hat{r}_\mi}
\def\MR{\hat{R}}
\def\MRe{\hat{R}_\me}
\def\MRi{\hat{R}_\mi}

\def\Mj{\hat{\jmath}}
\def\Mje{\hat{\jmath}_\me}
\def\Mji{\hat{\jmath}_\mi}

\def\Mq{\hat{q}}

\begin{document}

\title{Linear and non-linear integrate-and-fire neurons driven by\\ synaptic shot noise with reversal potentials}

\author{Magnus J. E. Richardson}

\affiliation{Warwick Mathematics Institute,  University of Warwick,\\ CV4 7AL, United Kingdom}
%\ead{magnus.richardson@warwick.ac.uk}

\begin{abstract}
The steady-state firing rate and firing-rate response of the leaky and
exponential integrate-and-fire models receiving synaptic shot noise
with excitatory and inhibitory reversal potentials is examined. For the particular case where the underlying synaptic conductances are exponentially distributed, it is shown that the master equation for a population of such model neurons can be reduced from an integro-differential form to a more tractable set of three differential equations. The system is nevertheless more challenging analytically than for current-based synapses: where possible analytical results are provided with an efficient numerical scheme and code provided for other quantities. The increased tractability of the framework developed supports an ongoing critical comparison between models in which synapses are treated with and without reversal potentials, such as recently in the context of networks with balanced excitatory and inhibitory conductances.\\[5mm]
\textbf{Reference: Physical Review E 109: 024407 (2024)}  
\end{abstract}

%\date

%\pacs{87.19.ll, 87.19.lc, 87.19.lq, 87.85.dm}
\maketitle

%%%%%%%%%%%%%%%%%%%%%%%%%%%%%%%%%%%%%%%%%%%%%%%%%%%%%%%%%%%%%%%%%%%
%%%%%%%%%%%%%%%%%%%%%%%%%%%%%%%%%%%%%%%%%%%%%%%%%%%%%%%%%%%%%%%%%%%
%%%%%%%%%%%%%%%%%%%%%%%%%%%%%%%%%%%%%%%%%%%%%%%%%%%%%%%%%%%%%%%%%%%
% INTRODUCTION
%%%%%%%%%%%%%%%%%%%%%%%%%%%%%%%%%%%%%%%%%%%%%%%%%%%%%%%%%%%%%%%%%%% 
%%%%%%%%%%%%%%%%%%%%%%%%%%%%%%%%%%%%%%%%%%%%%%%%%%%%%%%%%%%%%%%%%%%
%%%%%%%%%%%%%%%%%%%%%%%%%%%%%%%%%%%%%%%%%%%%%%%%%%%%%%%%%%%%%%%%%%%
\section{Introduction}
Neurons in active networks receive a barrage of competing excitatory and inhibitory synaptic input leading to a fluctuating membrane voltage and variability in the timing of outgoing spikes \cite{Shadlen1994}. Characterising how this incoming stochastic drive is integrated non-linearly and output spikes triggered is key to understanding single-neuron computation or how network states emerge and has been the subject of concerted theoretical effort for over half a century \cite{Stein1965,Johannesma1968,Knight1972,Ricciardi1977}. Over the years, increasing biophysical details have been incorporated into an analytical framework that includes processes in the synaptic and membrane-response components of neuronal integration and spike generation.

At the synaptic level, a common approach has been to approximate synaptic amplitudes as small so that, after a Gaussian approximation is made, a Fokker-Planck approach can be used to examine the neuronal or coupled network responses \cite{Brunel1999,Lindner2002,Gigante2007}. More recently, in part due to experimental evidence for long-tailed synaptic-amplitude distributions \cite{Song2005,Nobukawa2022} and effects of presynaptic synchrony \cite{DeWeese2006,Deniz2017}, there has been increasing interest in how finite-amplitude synaptic shot noise effects neuronal integration \cite{Knight2000,Burkitt2001,Kuhn2003,Helias2011,Richardson2010,Angulo-Garcia2017,Olmi2017,Droste2017,Richardson2018,Tamborrino2021}. Though the effect of synapses is often approximated as additive or current-based due to the reasonable desire for analytical tractability, they are more accurately implemented as conductances. Their stochastic activation therefore constitutes multiplicative noise due to the reversal-potential prefactor in the membrane current-balance equation. The aggregate conductance increase during strong presynaptic activity significantly affects the integrative properties of neurons \cite{Destexhe2003}, the Gaussianity of voltage fluctuations when coupled with finite-amplitude drive \cite{Richardson2005,Wolff2008,Privault2020}, the transmission of sensory signals \cite{Bauermann2019} and, importantly, qualitatively changes the nature of the modelled balanced state between excitation and inhibition \cite{Barta2021,Sanzeni2022}. 

At the membrane level, the majority of results for the stochastic integration of synaptic drive have been derived for neurons with a linear and ohmic response, specifically for the leaky integrate-and-fire (LIF) model - see \cite{Burkitt2006a,Burkitt2006b} for earlier reviews. However, from a theoretical reduction of Hodgkin-Huxley-type models, the current-voltage relationship was shown to be better captured by including an exponential non-linearity leading to the spike onset \cite{Fourcaud-Trocme2003}. It was subsequently shown experimentally that the resultant exponential integrate-and-fire (EIF) model provides a fairly accurate description of the integration properties of neocortical pyramidal cells \cite{Badel2008a} and fast-spiking interneurons \cite{Badel2008b}. 

Here, the framework for linear and non-linear integrate-and-fire neurons receiving exponentially distributed excitatory and inhibitory conductances is developed. The general framework previously introduced for population models \cite{Nykamp2000,Nykamp2001,Cai2006} with linear subthreshold behaviour is first followed. However, the choice of exponential conductance distributions (extending the approach for the LIF \cite{Richardson2010} and EIF \cite{Richardson2018} models driven by exponentially distributed additive shot noise) is shown to reduce the integro-differential master equation to purely differential form. It also allows for a more direct numerical solution, avoiding the interpolation and integrative steps needed for the more general gamma-distributed amplitudes treated in earlier work on such systems \cite{Nykamp2000}. After introducing the model at the level of stochastic single-neuron dynamics in Section II, the reduction of the master equation to a differential form is demonstrated in Section III. In section IV and V the framework is applied to the LIF and EIF models to examine the steady-state firing rate and firing-rate response. Though analytical solutions to the master equation when both inhibition and excitation are present were not found, an efficient numerical scheme for directly obtaining the steady state and rate response is developed with Julia code \cite{Bezanson2017} provided in the Supplemental Material \cite{MySupp}.

%%%%%%%%%%%%%%%%%%%%%%%%%%%%%%%%%%%%%%%%%%%%%%%%%%%%%%%%%%%%%%%%%%%
%%%%%%%%%%%%%%%%%%%%%%%%%%%%%%%%%%%%%%%%%%%%%%%%%%%%%%%%%%%%%%%%%%%
%%%%%%%%%%%%%%%%%%%%%%%%%%%%%%%%%%%%%%%%%%%%%%%%%%%%%%%%%%%%%%%%%%%
%%%%%%%%%%%%%%%%%%%%%%%%%%%%%%%%%%%%%%%%%%%%%%%%%%%%%%%%%%%%%%%%%%%
% Single neuron dynamics
%%%%%%%%%%%%%%%%%%%%%%%%%%%%%%%%%%%%%%%%%%%%%%%%%%%%%%%%%%%%%%%%%%% 
%%%%%%%%%%%%%%%%%%%%%%%%%%%%%%%%%%%%%%%%%%%%%%%%%%%%%%%%%%%%%%%%%%%
%%%%%%%%%%%%%%%%%%%%%%%%%%%%%%%%%%%%%%%%%%%%%%%%%%%%%%%%%%%%%%%%%%%
%%%%%%%%%%%%%%%%%%%%%%%%%%%%%%%%%%%%%%%%%%%%%%%%%%%%%%%%%%%%%%%%%%%
\section{Dynamics of a neuron} \vspace*{-2mm}
The neuron is isopotential with voltage $v(t)$ and receives stochastic drive from excitatory and inhibitory presynaptic populations. The rate of charging of the membrane capacitance is proportional to the voltage derivative, the intrinsic current-voltage relation of the neuron is modelled as without history, potentially non-linear and proportional to the function $f(v)$, and the stochastic synaptic current $s(v,t)$ is a function of both voltage and time as reversal potentials are included. Putting these terms together in the current-balance equation for the neuronal membrane yields a first-order non-linear differential equation driven by multiplicative noise
\be
\frac{dv}{dt}=f(v)+s(v,t). \label{dvdt}
\ee
The spike is implemented via the integrate-and-fire mechanism: if the voltage passes a threshold $\vth$ it is directly reset to a lower value $\vre$ and a spike is registered. 

\begin{figure}
\includegraphics[scale=0.625]{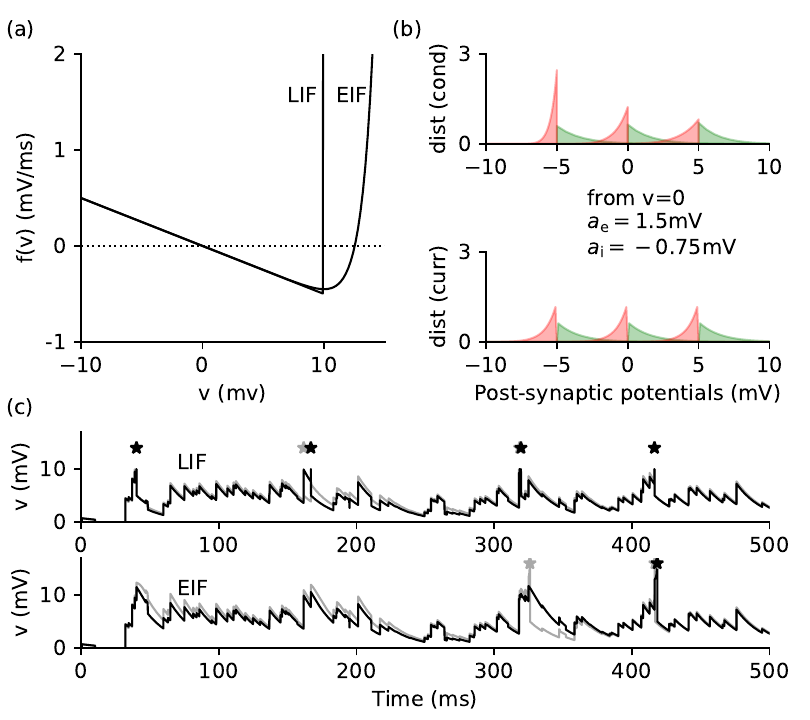}
\caption{Comparison of the Leaky and Exponential Integrate-and-Fire (LIF and EIF) models driven by exponentially distributed conductance or current-based synaptic shot noise. (a) Current-voltage relationship $f(v)$ normalised by the capacitance for the LIF and EIF models. (b) Excitatory (green) and inhibitory (red) synaptic-amplitude distributions as a function of three initial voltages ($-5,0,5$mV) for exponentially distributed synaptic conductances (upper panel, Eq. \ref{Be}) or currents (lower panel, Eq. \ref{Aexp}) with distributions having mean amplitudes matched at $v\!=\!0$. (c) Voltage dynamics for the LIF (upper panel) and EIF (lower panel). Implementations with conductance-based (Eq. \ref{Scond}, black lines) and current-based (Eq. \ref{Scurr}, grey lines) drive for the distributions in panel (b) exhibit clear differences in membrane voltage and firing times. Parameters: $\tau\!=\!20$ms, $\dT\!=\!1$mV, $\vT\!=\!10$mV, $\vre\!=\!5$mV, $\vth\!=\!10$mV for the LIF or $\vth\!=\!20$mV for the EIF; reversal potentials $\epse=60$mV,~$\epsi=-10$mV; mean synaptic amplitudes as marked; steady-state presynaptic rates for panel (c) were $\SRe\!=\!0.2$kHz and $\SRi\!=\!0.02$kHz. Code used to generate the figure is provided in the Supplemental Material \cite{MySupp}.}
\label{fig1}
\end{figure}

Two different current-voltage relationships are considered in this paper. The first is a leaky integrator
\be
f(v)=-\frac{v}{\tau} \label{FLIF}
\ee
where $\tau$ is the membrane time constant. The second forcing term considered is that of the EIF \cite{Fourcaud-Trocme2003} which includes the non-linear effect of the spike onset
\be
f(v)=\frac{1}{\tau}\l(\dT e^{(v-\vT)/\dT}- v\r). \label{FEIF}
\ee
As well as a stable resting voltage near $v\!=\!0$, the model features an intrinsic spiking mechanism via an unstable fixed point. Above this value, the voltage diverges until it hits the threshold and is then reset. The quantity $\vT$ is the voltage at which the forcing term is at its minimum whereas $\dT$ parameterises how the exponential non-linearity grows. These two $f(v)$ choices, corresponding to the LIF and EIF models, are plotted in Fig. \ref{fig1}a.

%%%%%%%%%%%%%%%%%%%%%%%%%%%%%%%%%%%%%%%%%%%%%%%%%%%%%%%%%%%%%%%%%%%
%%%%%%%%%%%%%%%%%%%%%%%%%%%%%%%%%%%%%%%%%%%%%%%%%%%%%%%%%%%%%%%%%%%
% Figure 1. Forcing term and the dynamics
%%%%%%%%%%%%%%%%%%%%%%%%%%%%%%%%%%%%%%%%%%%%%%%%%%%%%%%%%%%%%%%%%%%
%%%%%%%%%%%%%%%%%%%%%%%%%%%%%%%%%%%%%%%%%%%%%%%%%%%%%%%%%%%%%%%%%%%
%%%%%%%%%%%%%%%%%%%%%%%%%%%%%%%%%%%%%%%%%%%%%%%%%%%%%%%%%%%%%%%%%%%
%%%%%%%%%%%%%%%%%%%%%%%%%%%%%%%%%%%%%%%%%%%%%%%%%%%%%%%%%%%%%%%%%%%
  
%%%%%%%%%%%%%%%%%%%%%%%%%%%%%%%%%%%%%%%%%%%%%%%%%%%%%%%%%%%%%%%%%%%
%%%%%%%%%%%%%%%%%%%%%%%%%%%%%%%%%%%%%%%%%%%%%%%%%%%%%%%%%%%%%%%%%%%
% Conductance-based synaptic shot noise
%%%%%%%%%%%%%%%%%%%%%%%%%%%%%%%%%%%%%%%%%%%%%%%%%%%%%%%%%%%%%%%%%%%
%%%%%%%%%%%%%%%%%%%%%%%%%%%%%%%%%%%%%%%%%%%%%%%%%%%%%%%%%%%%%%%%%%%
\subsection*{Shot noise with reversal potentials} \vspace*{-2mm}
The second term on the right-hand-side of Eq. (\ref{dvdt}) is now examined in detail. In this paper, a synaptic input is approximated as being unfiltered and implemented as a conductance impulse having the effect of immediately increasing (excitation) or decreasing (inhibition) the membrane potential by a voltage-dependent amplitude. In the context of active presynaptic populations, the barrage of excitatory and inhibitory input is modelled by time-dependent Poissonian rates $\Re(t)$ and $\Ri(t)$ such that
\be
\hspace{-3mm}s(v,t)\!=\!(\epse\!-\!v)\!\sum_{\{t^\me_k\}}h^\me_{k}\delta(t\!-\!t^\me_{k})\!+\!(\epsi\!-\!v)\!\sum_{\{t^\mi_k\}}h^\mi_{k}\delta(t\!-\!t^\mi_{k}) \label{Scond}
\ee
where $\epse$ and $\epsi$ are the reversal potentials and $\{t^\me_k\}$, $\{t^\mi_k\}$ the set of presynaptic-pulse arrival times. The quantities $h^\me_k$ and $h^\mi_k$ are unitless (scaled conductance) impulses drawn from a biophysically plausible distribution. In the context of the fast-synapse limit used here, these quantities can be thought of as being equal to the time-integral of a fast synaptic-conductance waveform divided by the membrane capacitance and therefore directly proportional to the distribution of synaptic conductance strengths

Care is needed in the interpretation of the effect of $s(v,t)$ on the voltage dynamics in Eq. (\ref{dvdt}) due to the delta functions having a voltage-dependent prefactor. Using an isolated excitatory pulse as an example
\be
\frac{dv}{dt}=f(v)+(\epse-v)h\delta(t),
\ee
both sides are first divided by $(\epse\!-\!v)$ and then integrated over a short time window that includes the delta pulse. The term including $f(v)$ will vanish with the size of this time window leaving the solution
\be
\log\l(\frac{\epse-w}{\epse-v}\r)=h \label{hvw}
\ee
where $w$ was the voltage before and $v\!>\!w$ is the voltage
after the pulse. This equation can be re-arranged to give the voltage jump from its initial value $w$ as
\be
v-w=(\epse-w)b \label{bform}
\ee
where $b\!=\!1\!-\!e^{-h}$ is a convenient measure of the synaptic amplitude \cite{Nykamp2000} and is bounded between $0$ and $1$. The amplitudes of the synaptic impulses in Eq. (\ref{Scond}) are stochastic and drawn, using excitation as an example, from a distribution $H_\me(h)$. Of interest, in the population-level description to be considered later, will be the fraction of synaptic impulses that bring the voltage above some value $v$ from a lower value $w$. This tail distribution can be written as
\be
T_\me(v,w)=\int_{h_{vw}}^\infty dh H_\me(h) \label{TailDist}
\ee
where the lower bound $h_{vw}$ is the value given in Eq. (\ref{hvw}) and ensures that only jumps taking the voltage above $v$ are included.

%%%%%%%%%%%%%%%%%%%%%%%%%%%%%%%%%%%%%%%%%%%%%%%%%%%%%%%%%%%%%%%%%%%
%%%%%%%%%%%%%%%%%%%%%%%%%%%%%%%%%%%%%%%%%%%%%%%%%%%%%%%%%%%%%%%%%%%
% Exponential choice for the conductance distribution
%%%%%%%%%%%%%%%%%%%%%%%%%%%%%%%%%%%%%%%%%%%%%%%%%%%%%%%%%%%%%%%%%%%
%%%%%%%%%%%%%%%%%%%%%%%%%%%%%%%%%%%%%%%%%%%%%%%%%%%%%%%%%%%%%%%%%%%
\subsection*{Exponentially distributed synaptic conductances} \vspace*{-2mm}
A biophysically plausible exponential amplitude distribution of conductance impulses is chosen here. As will be seen later, this special case of the more general gamma distribution used in reference \cite{Nykamp2000} allows for a considerable simplification of both the analytical description of the master equation and its numerical solution at the population level. Using excitation as an example, the distribution is written
\be
H_\me(h)=\theta(h)\frac{e^{-h/h_\me}}{h_\me} \label{hexp}
\ee
where $h_\me$ is the mean conductance amplitude. As introduced earlier in Eq. (\ref{bform}), the voltage increase can be conveniently expressed in a transformed variable $b\!=\!1\!-\!e^{-h}$ where $b\!=\!0$ corresponds to a negligibly small input and $b\!=\!1$ an input sufficiently strong to place the voltage right at the excitatory reversal potential $\epse$. An exponential distribution for $h$ implies a special case of the beta distribution for $b$
\be
B_\me(b)\!=\!\betae(1\!-\!b)^{\betae-1} \label{Be}
\ee
where $\betae\!=\!1/\hme$ and the mean of $b$ is $\bme\!=\!1/(\betae\!+\!1)$. The distribution parameters $\hme$ or $\betae$ can therefore be related to the mean synaptic amplitude from rest $\ame$ via Eq. (\ref{bform}) which is
\be
\ame=\epse\bme.
\ee
This gives the typical jump size from a voltage $w$ as $\ame(1\!-\!w/\epse)$. Note that with this definition the limit $\epse\to\infty$ with $\ame$ held constant recovers a current-based implementation of the synaptic drive (see also Eq. \ref{Scurr} of the Appendix). Examples of these results, and the equivalent for inhibition, are presented in Fig. \ref{fig1}b (upper panel) where the amplitude distributions from three different initial voltages are shown. Note that the inhibitory amplitude distribution becomes markedly sharper as the inhibitory reversal potential is neared. As a reference, the distributions for exponentially distributed current-based synaptic drive (see Appendix Eqs. \ref{Scurr} and \ref{Aexp}) are provided in Fig. \ref{fig1}b lower panel showing the expected constancy at different initial voltages. Finally, in Fig. \ref{fig1}c example voltage dynamics (Eq. \ref{dvdt}) for the LIF and EIF are provided in the upper and lower panels. The same patterned stochastic inputs with amplitudes matched at rest are provided to conductance-based (Eq. \ref{Scond}) and current-based (Appendix Eq. \ref{Scurr}) implementations of the drive demonstrating distinct responses.

%%%%%%%%%%%%%%%%%%%%%%%%%%%%%%%%%%%%%%%%%%%%%%%%%%%%%%%%%%%%%%%%%%%
%%%%%%%%%%%%%%%%%%%%%%%%%%%%%%%%%%%%%%%%%%%%%%%%%%%%%%%%%%%%%%%%%%%
%%%%%%%%%%%%%%%%%%%%%%%%%%%%%%%%%%%%%%%%%%%%%%%%%%%%%%%%%%%%%%%%%%%
%%%%%%%%%%%%%%%%%%%%%%%%%%%%%%%%%%%%%%%%%%%%%%%%%%%%%%%%%%%%%%%%%%%
% Population-density description
%%%%%%%%%%%%%%%%%%%%%%%%%%%%%%%%%%%%%%%%%%%%%%%%%%%%%%%%%%%%%%%%%%%
%%%%%%%%%%%%%%%%%%%%%%%%%%%%%%%%%%%%%%%%%%%%%%%%%%%%%%%%%%%%%%%%%%%
%%%%%%%%%%%%%%%%%%%%%%%%%%%%%%%%%%%%%%%%%%%%%%%%%%%%%%%%%%%%%%%%%%%
%%%%%%%%%%%%%%%%%%%%%%%%%%%%%%%%%%%%%%%%%%%%%%%%%%%%%%%%%%%%%%%%%%%
\section{Population dynamics} \vspace*{-2mm}
The voltage trajectories of a population of neurons, each obeying Eq. (\ref{dvdt}) with an uncorrelated but statistically identical realisation of the stochastic drive in Eq. (\ref{Scond}), can be described by a probability density $P(v,t)$. The stochastic single neuronal dynamics allow for the construction of a master equation that describes the deterministic dynamics of the ensemble at the population level. As well as the probability density, it is convenient to consider the probability flux $J(v,t)$. This describes the flow-rate of trajectories passing a particular voltage. Note that the flux at threshold $J(\vth)$ is equal to the instantaneous spike-rate
$r(t)$ of the population with the flow then reinserted at the reset $\vre$. These quantities are connected by a continuity equation 
\be
\frac{\p P}{\p t}+\frac{\p J}{\p
  v}=r(t)\l(\delta(v-\vre)-\delta(v-\vth)\r) \label{cont}
\ee
which is statement of conservation of total density. The flux $J$ can be resolved into a deterministic contribution equal to the forcing term multiplied by the density $fP$ and two terms $\Je\!>\!0$ and $\Ji\!<\!0$ coming from the stochastic excitatory and inhibitory synaptic events in the $s(v,t)$ term of Eq. (\ref{dvdt}). This gives the flux equation
\be
J=fP+\Je+\Ji. \label{flux}
\ee
The synaptic fluxes $\Je$, $\Ji$ for conductance-based synaptic drive $s(v,t)$ can be straightforwardly derived from the amplitude distributions considered in the previous section. For example, the excitatory flux across a voltage $v$ is the excitatory presynaptic rate times the fraction of amplitudes that bring the neuron from any lower voltage $w$ to a voltage greater than $v$. Hence,
\be
\Je(v,t)=\Re\int_{-\infty}^vdwP(w)T_\me(v,w) \label{JeT}
\ee
where $T_\me(v,w)$ is the tail distribution given in Eq. (\ref{TailDist}). A similar form is derivable for inhibition but will be negative. Equations (\ref{cont}-\ref{JeT}) constitute the master equation for fast synaptic shot noise implemented with reversal potentials and having a general distribution for the synaptic-conductance amplitudes. Such coupled integro-differential equations sets are generally difficult to treat analytically. In the next section it is shown that the description simplifies considerably when the underlying distribution of synaptic conductances is exponential.

%%%%%%%%%%%%%%%%%%%%%%%%%%%%%%%%%%%%%%%%%%%%%%%%%%%%%%%%%%%%%%%%%%%
%%%%%%%%%%%%%%%%%%%%%%%%%%%%%%%%%%%%%%%%%%%%%%%%%%%%%%%%%%%%%%%%%%%
% Exponential choice for the conductance distribution
%%%%%%%%%%%%%%%%%%%%%%%%%%%%%%%%%%%%%%%%%%%%%%%%%%%%%%%%%%%%%%%%%%%
%%%%%%%%%%%%%%%%%%%%%%%%%%%%%%%%%%%%%%%%%%%%%%%%%%%%%%%%%%%%%%%%%%%
\subsection*{Differential form of the master equation} \vspace*{-2mm} 
The exponential form for $H_\me$ given above (Eq. \ref{hexp}) is now substituted into the tail equation for the distribution $T_\me(v,w)$ and integrated to give
\be
\Je(v,t)=\Re\int_{-\infty}^vdwP(w)e^{-h_{vw}/h_\me}. \label{Jeh}
\ee
Substituting in for $h_{vw}$ given in Eq. (\ref{hvw}) yields 
\be
\Je(v,t)=\Re\int_{-\infty}^vdwP(w)\l(\frac{\epse-w}{\epse-v}\r)^{\betae} \label{Jeb}
\ee
where $\betae\!=\!1/h_\me$. It can be noted that this is a special case of the gamma-distributed conductance amplitudes considered previously \cite{Nykamp2000}. Following the approach in reference \cite{Richardson2010}, the derivative with respect to voltage is taken and the resulting integral in one of the terms identified as being proportional to $\Je$. This results in differential equations for the excitatory and inhibitory synaptic fluxes that take similar forms
\be
\hspace*{-8mm}\frac{\p \Je}{\p v}+\frac{\beta_e\Je}{\epse-v}=R_eP &\mbox{ and }& \frac{\p \Ji}{\p v}+\frac{\beta_i\Ji}{\epsi-v}=R_iP. \label{GJeJi}
\ee
These two synaptic flux equations together with the continuity
(\ref{cont}) and flux (\ref{flux}) equations describe the
dynamics of an ensemble of neurons subject to exponentially distributed conductance-based shot noise.

%%%%%%%%%%%%%%%%%%%%%%%%%%%%%%%%%%%%%%%%%%%%%%%%%%%%%%%%%%%%%%%%%%%
%%%%%%%%%%%%%%%%%%%%%%%%%%%%%%%%%%%%%%%%%%%%%%%%%%%%%%%%%%%%%%%%%%%
% FIRING-RATE RESPONSE FOR CONDUCTANCE-BASED SHOT NOISE
%%%%%%%%%%%%%%%%%%%%%%%%%%%%%%%%%%%%%%%%%%%%%%%%%%%%%%%%%%%%%%%%%%% 
%%%%%%%%%%%%%%%%%%%%%%%%%%%%%%%%%%%%%%%%%%%%%%%%%%%%%%%%%%%%%%%%%%%
\subsection*{Steady-state rate and firing-rate response} \vspace*{-2mm} 
The master equation describes the full dynamics with arbitrarily strong modulations of the incoming synaptic rates $\Re(t)$ and $\Ri(t)$. A full solution of the system appears difficult to obtain, given the complexity of the time-voltage partial-differential equation set and threshold-reset boundary conditions. However, simpler quantities such as the steady-state rate and firing-rate response - the response to weak modulations of the incoming rates - nevertheless provide important information on the behaviour at the neuronal-population level and are central quantities needed for network stability and emergent oscillations \cite{Brunel1999}. Using modulated excitation as an example, the incoming rate can be written in complex form as
\be
\Re(t)=\SRe+\MRe e^{i\w t}. 
\ee
where $\SRe$ is the steady-state value, $\MRe$ the (potentially complex) amplitude of the modulation and $\w$ its angular frequency. The modulatory amplitude is considered sufficiently weak such that evoked modulations in all down-stream variables take (using the excitatory flux as an example) the form
\be
\Je(t)=\SJe+\MJe e^{i\w t}.
\ee
These expansions can be substituted into the master equation defined through Eqs. (\ref{cont},\ref{flux},\ref{GJeJi}) with the resulting steady-state quantities providing a self-contained set of equations and the modulations providing a second set of equations. The master equations in the steady-state is 
\be
&&\frac{d\SJ}{dv}=\Sr\l(\delta(v-\vre)-\delta(v-\vth)\r) \label{dJdv0con}\\
&&\frac{d \SJe}{d
  v}+\frac{\betae\SJe}{\epse-v}=\SRe\SP\label{dJedv0con}. \\
&&\frac{d
  \SJi}{dv}+\frac{\betai\SJi}{\epsi-v}=\SRi\SP \label{dJidv0con} 
\ee
with the steady-state flux equation $\SJ\!=\!f\SP+\SJe+\SJi$. Note that the first equation implies $\SJ=\Sr\theta(v-\vre)$ for $v<\vth$. Next, at the level of a weak excitatory and inhibitory modulation of the master equation a similar-looking set of differential equations is found 
\be
&&i\w\MP+\frac{d\MJ}{dv}=\Mr\l(\delta(v-\vre) - \delta(v-\vth)\r) \label{dJdv1con}\\
&&\frac{d \MJe}{d
  v}+\frac{\betae\MJe}{\epse-v}=\SRe\MP+\MRe\SP\label{dJedv1con}. \\
&&\frac{d \MJi}{dv}+\frac{\betai\MJi}{\epsi-v}=\SRi\MP +\MRi\SP\label{dJidv1con} 
\ee
where now $\MJ\!=\!f\MP+\MJe+\MJi$. For these quantities, the master equation has therefore been reduced to a set of coupled ordinary differential equations for which there is some hope of analytical solution or at least an efficient numerical scheme. Before treating the equations, a general statement about the limit of their behaviour at high frequency modulation is first provided.

%%%%%%%%%%%%%%%%%%%%%%%%%%%%%%%%%%%%%%%%%%%%%%%%%%%%%%%%%%%%%%%%%%%
%%%%%%%%%%%%%%%%%%%%%%%%%%%%%%%%%%%%%%%%%%%%%%%%%%%%%%%%%%%%%%%%%%%
% Synaptic fluxes at high frequency
%%%%%%%%%%%%%%%%%%%%%%%%%%%%%%%%%%%%%%%%%%%%%%%%%%%%%%%%%%%%%%%%%%%
%%%%%%%%%%%%%%%%%%%%%%%%%%%%%%%%%%%%%%%%%%%%%%%%%%%%%%%%%%%%%%%%%%%
\subsection*{Synaptic fluxes at high frequency} \vspace*{-2mm} 
For the analysis of the high-frequency asymptotics, it is useful to consider the behaviour of the synaptic fluxes in that regime. First, it can be noted that because of the prefactor $i\w$ in the 
continuity equation (\ref{dJdv1con}), the modulated probability density $\MP$ will decay with frequency. This allows for the following observation, again using excitation as an example: expanding the integral (Eq. \ref{Jeb}) in terms of the modulated components gives
\be
\hspace*{-9mm}\MJe\!&\!\!=\!\!&\!\MRe\!\int_{\epsi}^v\!\!du\SP(u)\!\l(\frac{\epse\!-\!v}{\epse\!-\!u}\r)^{\betae}\!\!\!\!\!+\!\SRe\!\int_{\epsi}^v\!\!du\MP(u)\!\l(\frac{\epse\!-\!v}{\epse\!-\!u}\r)^{\betae}\!\!\!\!\!\!. \label{expJe}
\ee
As $\MP\!\to\!0$ at high frequency, the first term becomes increasingly dominant. It can be noted that this term is simply the steady-state excitatory synaptic flux multiplied by the ratio $\MRe/\SRe$. This argument follows for both excitatory and inhibitory modulation and so, for increasing high modulation frequencies, the modulated synaptic fluxes can be written in terms of their steady-state values
\be
\MJe\to\frac{\MRe}{\SRe}\SJe &\mbox{ and }& \MJi\to\frac{\MRi}{\SRi}\SJi. \label{JeJiAsym}
\ee
These results allow for the high-frequency firing-rate response to be related to steady-state quantities significantly simplifying the asymptotic analysis.

%%%%%%%%%%%%%%%%%%%%%%%%%%%%%%%%%%%%%%%%%%%%%%%%%%%%%%%%%%%%%%%%%%%
%%%%%%%%%%%%%%%%%%%%%%%%%%%%%%%%%%%%%%%%%%%%%%%%%%%%%%%%%%%%%%%%%%%
% Figure 2. LIF model
%%%%%%%%%%%%%%%%%%%%%%%%%%%%%%%%%%%%%%%%%%%%%%%%%%%%%%%%%%%%%%%%%%%
%%%%%%%%%%%%%%%%%%%%%%%%%%%%%%%%%%%%%%%%%%%%%%%%%%%%%%%%%%%%%%%%%%%
\begin{figure*}
\centerline{\includegraphics[scale=0.625]{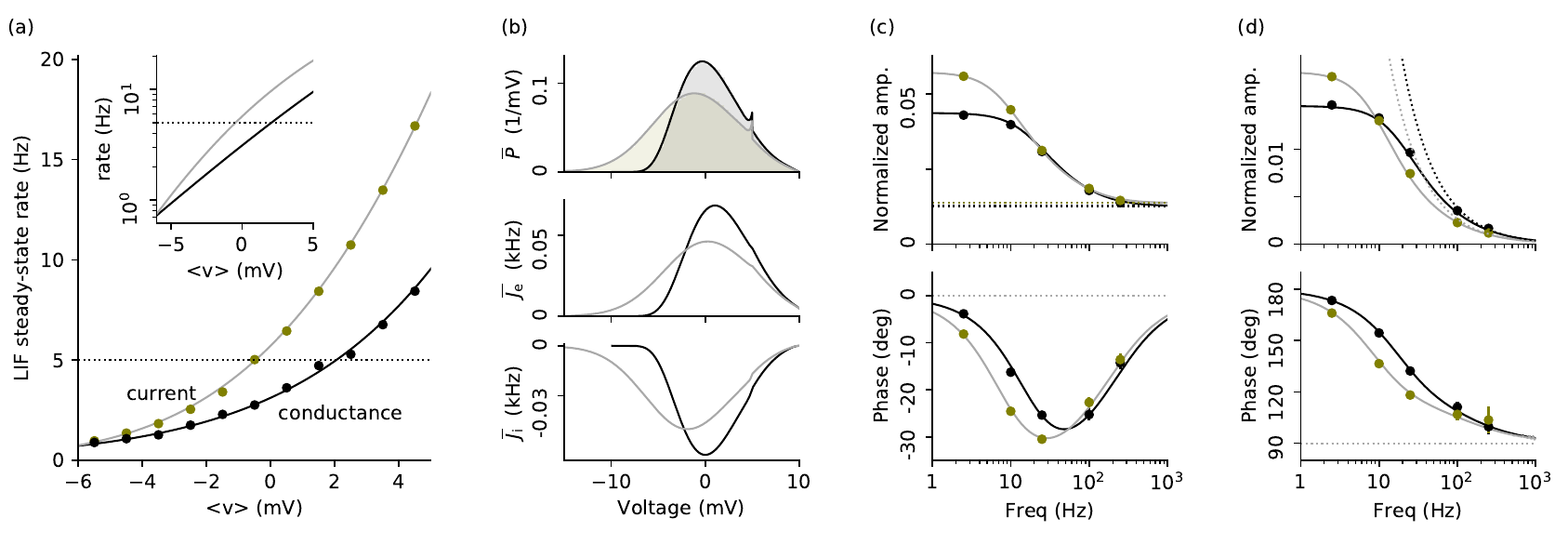}}
\caption{Leaky I\&F neuron driven by conductance or current-based shot noise with excitatory and inhibitory reversal potentials. (a) Steady-state outgoing firing rate with the incoming presynaptic rates $(\SRe,~\SRi)$ the same for the conductance and current-based cases. The presynaptic rates were parameterised, for convenience, from the current-based equations (see Eq. \ref{Curmusig}) for the voltage mean $\Ex{v}$ with the standard deviation fixed at $\sigma\!=\!5$mV. Note the near-exponential rise for the conductance case seen in the inset. (b) Probability density together with excitatory and inhibitory fluxes for neurons firing at $5$Hz (intersections with dotted line in panel a). The presynaptic rates $(\SRe,~\SRi)$ were $(0.393,0.650)$kHz and $(0.365,0.762)$kHz for conductance and current-based cases, respectively. (c) Amplitude and phase of the firing-rate response to modulation of the presynaptic excitatory rate with asymptotics shown. (d) Same, but for the case of modulation of the inhibitory presynaptic rate. For panels (c-d) the presynaptic rates from panel (b) were used (corresponding to $\Sr\!=\!5$Hz) with all other parameters for the LIF provided in caption to Fig. 1. Simulational results are provided by symbols in panels (a,c,d) and histograms in upper panel (b). All code used to generate the figure is provided in the Supplemental Material \cite{MySupp}.}
\label{fig2}
\end{figure*}
%%%%%%%%%%%%%%%%%%%%%%%%%%%%%%%%%%%%%%%%%%%%%%%%%%%%%%%%%%%%%%%%%%%
%%%%%%%%%%%%%%%%%%%%%%%%%%%%%%%%%%%%%%%%%%%%%%%%%%%%%%%%%%%%%%%%%%%
  
%%%%%%%%%%%%%%%%%%%%%%%%%%%%%%%%%%%%%%%%%%%%%%%%%%%%%%%%%%%%%%%%%%%
%%%%%%%%%%%%%%%%%%%%%%%%%%%%%%%%%%%%%%%%%%%%%%%%%%%%%%%%%%%%%%%%%%%
%%%%%%%%%%%%%%%%%%%%%%%%%%%%%%%%%%%%%%%%%%%%%%%%%%%%%%%%%%%%%%%%%%%
%%%%%%%%%%%%%%%%%%%%%%%%%%%%%%%%%%%%%%%%%%%%%%%%%%%%%%%%%%%%%%%%%%%
% LEAKY INTEGRATE-AND-FIRE MODEL
%%%%%%%%%%%%%%%%%%%%%%%%%%%%%%%%%%%%%%%%%%%%%%%%%%%%%%%%%%%%%%%%%%% 
%%%%%%%%%%%%%%%%%%%%%%%%%%%%%%%%%%%%%%%%%%%%%%%%%%%%%%%%%%%%%%%%%%%
%%%%%%%%%%%%%%%%%%%%%%%%%%%%%%%%%%%%%%%%%%%%%%%%%%%%%%%%%%%%%%%%%%%
%%%%%%%%%%%%%%%%%%%%%%%%%%%%%%%%%%%%%%%%%%%%%%%%%%%%%%%%%%%%%%%%%%%
\section{Leaky Integrate \& Fire model} \vspace*{-2mm} 
For the LIF model, the current-voltage relation $f(v)$ is linear (Eq. \ref{FLIF}, Fig. \ref{fig1}a) and the threshold for action potential generation $\vth$ is at the beginning of the spike with an instantaneous reset to $\vre$. The threshold and reset are both considered to be above the resting potential, with the former criterion ensuring that the neuron is in the fluctuation-driven firing regime.

%%%%%%%%%%%%%%%%%%%%%%%%%%%%%%%%%%%%%%%%%%%%%%%%%%%%%%%%%%%%%%%%%%%
% Boundary conditions
%%%%%%%%%%%%%%%%%%%%%%%%%%%%%%%%%%%%%%%%%%%%%%%%%%%%%%%%%%%%%%%%%%%
\bigskip \ni\textit{Boundary conditions.} In the fluctuation-driven regime, the threshold can only be crossed by an excitatory synaptic event so the firing rate is identical to the excitatory flux at threshold:
\be
\hspace*{-7mm} r(t)\!=\!J(\vth,t)\!=\!\Je(\vth,t) &\mbox{ implying }& P(\vth,t)\!=\!0. \label{rJeLIF}
\ee
The second result - zero probability density at threshold - comes from the flux equation (\ref{flux}) and the equality of the total and excitatory steady-state fluxes combined with zero inhibitory flux at threshold (there are no neurons with $v\!>\!\vth$). This requires $P(\vth)\!=\!0$ given that the forcing term $f(\vth)$ is non-zero at threshold.

%%%%%%%%%%%%%%%%%%%%%%%%%%%%%%%%%%%%%%%%%%%%%%%%%%%%%%%%%%%%%%%%%%%
%%%%%%%%%%%%%%%%%%%%%%%%%%%%%%%%%%%%%%%%%%%%%%%%%%%%%%%%%%%%%%%%%%%
% Steady-state for the LIF with conductance-based shot noise
%%%%%%%%%%%%%%%%%%%%%%%%%%%%%%%%%%%%%%%%%%%%%%%%%%%%%%%%%%%%%%%%%%%
%%%%%%%%%%%%%%%%%%%%%%%%%%%%%%%%%%%%%%%%%%%%%%%%%%%%%%%%%%%%%%%%%%%
\subsection*{LIF steady-state rate}\vspace*{-2mm}
For the LIF driven by current-based shot noise \cite{Richardson2010} it was possible to solve
the master equation using a bilateral Laplace transform. This approach does not appear practical for combined excitatory and inhibitory shot noise with reversal potentials due to the additional voltage dependencies in the flux equations (\ref{GJeJi}) compared to those for current-based drive (see Appendix Eq. \ref{IJeJi}). However, an efficient numerical scheme developed for additive Gaussian \cite{Richardson2007} and additive shot-noise \cite{Richardson2010} can be extended to account for the reversal potentials. The method works by integrating Eqs. (\ref{dJdv0con}-\ref{dJidv0con}) in the direction of convergence in two domains: 
\be
[\epsi\xrightarrow{(i)} 0]~[0\xleftarrow{(ii)}\vth]. \label{LIFdomains}
\ee
The only inhomogeneous term in the equation set is proportional to $\Sr$ (in Eq. \ref{dJdv0con}). It must be, therefore, that all quantities are proportional to $\Sr$ and so this factor can be scaled out. Using lower-case letters for the scaled versions, the flux is trivially given as $\Sj=\theta(v\!-\!\vre)$ from the solution of Eq. (\ref{dJdv0con}). This leaves the remaining pair of synaptic fluxes $(\Sje,\Sji)$ to describe the system where the scaled probability density can be written in terms of these quantities by re-arranging the flux equation $\Sp\!=\!(\Sj-\Sje-\Sji)/f$. For domain (ii) the equations (\ref{dJedv0con},\ref{dJidv0con}) with threshold conditions $(1,0)$ are integrated downwards to $v\!=\!0$. For domain (i) the equations are integrated up to $v\!=\!0$ starting from just above the inhibitory reversal potential with initial conditions $(0,-1)$. The excitatory flux is then matched on either side of the origin by scaling the solutions in domain (i). Finally, the unknown steady-state rate $\Sr$ is recovered by the normalisation condition on the probability density $\SP(v)$ via $\Sr\!\int\!\Sp(v) dv\!=\!1$. 

The steady-state rate for excitatory and inhibitory conductance-based synaptic shot noise is given for the LIF model in Fig. \ref{fig2}a as a function of the subthreshold mean voltage. The inset features a semilog plot on the $y$-axis shows that the rate is close to exponential with the mean voltage. The behaviour is also compared to that for additive shot noise (see Appendix B). In Fig. \ref{fig2}b the steady-state probability density and synaptic fluxes are shown for the case where $\Sr\!=\!5$Hz. Note that the distribution of the current-based case is curtailed above the reversal potential for inhibition $\epsi$.

%%%%%%%%%%%%%%%%%%%%%%%%%%%%%%%%%%%%%%%%%%%%%%%%%%%%%%%%%%%%%%%%%%%
%%%%%%%%%%%%%%%%%%%%%%%%%%%%%%%%%%%%%%%%%%%%%%%%%%%%%%%%%%%%%%%%%%%
% LIF modulatory response
%%%%%%%%%%%%%%%%%%%%%%%%%%%%%%%%%%%%%%%%%%%%%%%%%%%%%%%%%%%%%%%%%%%
%%%%%%%%%%%%%%%%%%%%%%%%%%%%%%%%%%%%%%%%%%%%%%%%%%%%%%%%%%%%%%%%%%%
\subsection*{LIF firing-rate response}\vspace*{-2mm}
It was not obvious how an analytical solution to Eqs. (\ref{dJdv1con}-\ref{dJidv1con}) for the modulated rate can be obtained for combined excitatory and inhibitory conductance-based shot noise; however, it is again straightforward to generalise the numerical approach. At the modulated level, there are two inhomogeneous terms, proportional to $\Mr$ and either $\MRe$ or $\MRi$ in the equation set (taking either excitatory or inhibitory modulation separately). The method is therefore more involved than for the steady-state case and can be found in full in the Appendix. Examples of the firing-rate response are provided in Fig. \ref{fig2}c for modulated excitation and in Fig. \ref{fig2}d for modulated inhibition. They are compared with previously derived results for current-based shot noise and shown to be broadly similar, notably in their high-frequency asymptotics. Though it was not possible to solve the equations analytically, the asymptotics for the firing-rate response for the LIF model can be derived.

%%%%%%%%%%%%%%%%%%%%%%%%%%%%%%%%%%%%%%%%%%%%%%%%%%%%%%%%%%%%%%%%%%%
% Excitatory modulation for the LIF
%%%%%%%%%%%%%%%%%%%%%%%%%%%%%%%%%%%%%%%%%%%%%%%%%%%%%%%%%%%%%%%%%%%
\bigskip\ni\textit{Excitatory modulation at high frequencies.} For the LIF, the modulated firing rate is  given by the modulated excitatory flux at threshold $\MJe(\vth)$ as stated in Eq. (\ref{rJeLIF}). This quantity in turn becomes increasingly proportional to the steady-state excitatory flux as the modulating frequency increases (see Eq. \ref{JeJiAsym}). Applying this gives the result 
\be
\Mr\sim\frac{\MRe}{\SRe}\Sr
\ee
and so the response to modulation of the presynaptic excitatory rate tends to a constant in the limit of high frequency with zero phase lag. An example of this asymptotic behaviour is provided in Fig. \ref{fig2}c and compared to the current-based case that has an identical form.

%%%%%%%%%%%%%%%%%%%%%%%%%%%%%%%%%%%%%%%%%%%%%%%%%%%%%%%%%%%%%%%%%%%
% Inhibitory modulation for the LIF
%%%%%%%%%%%%%%%%%%%%%%%%%%%%%%%%%%%%%%%%%%%%%%%%%%%%%%%%%%%%%%%%%%%
\bigskip\ni\textit{Inhibitory modulation at high frequencies.} Though it is the incoming inhibitory rate that is modulated, the outgoing firing rate is still given by the modulated excitatory flux but with the first term in the expansion Eq. (\ref{expJe}) absent as $\MRe\!=\!0$. The modulated firing rate therefore reduces to just the second term and so 
\be
\Mr=\SRe\int_{\epsi}^{\vth}du\l(\frac{\epse\!-\!v_{th}}{\epse\!-\!u}\r)^{\betae}\MP(u) \label{hri1}
\ee
where it remains only to approximate $\MP(u)$ at high frequency. The argument is as follows: in Eq. (\ref{dJidv1con}) there is an inhomogeneous term that is proportional to the steady-state density $\SP$ that does not decay with increasing frequency. Driven by this term, inhibitory flux is therefore dominant over the $f\MP$ and $\MJe$ terms in the modulated flux equation, so that $\MJ\!\simeq\!\MJi$ at high frequencies. Inserting this into equation (\ref{dJdv1con}) gives $\MP=-(d \MJi/dv)/i\w$ and then, using the result from Eq. (\ref{JeJiAsym}) that $\MJi\!\simeq\!\SJi\MRi/\SRi$ at high frequencies allows for substitution into Eq. (\ref{hri1}). Following a final integration-by-parts step gives the modulated rate in terms of the steady-state inhibitory flux
\be
\Mr\sim\MRi\frac{\SRe}{\SRi}\frac{\betae}{i\w}\int_{\epsi}^{v_{th}}du\l(\frac{\epse\!-\!v_{th}}{\epse\!-\!u}\r)^{\beta_e}\frac{\SJi(u)}{\epse-u}.
\ee 
It was not obvious how to further reduce this equation. Nevertheless, it shows that the high-frequency asymptotics decay with the reciprocal of frequency and with a phase of $90^\circ$ by virtue of the $1/i$ term and that the steady-state inhibitory flux $\SJi$ is negative. An example of this result is provided in Fig. \ref{fig2}d and compared to the result for current-based inhibitory shot noise which shares the same asymptotic behaviour.

%%%%%%%%%%%%%%%%%%%%%%%%%%%%%%%%%%%%%%%%%%%%%%%%%%%%%%%%%%%%%%%%%%%
%%%%%%%%%%%%%%%%%%%%%%%%%%%%%%%%%%%%%%%%%%%%%%%%%%%%%%%%%%%%%%%%%%%
% Figure 3. EIF model
%%%%%%%%%%%%%%%%%%%%%%%%%%%%%%%%%%%%%%%%%%%%%%%%%%%%%%%%%%%%%%%%%%%
%%%%%%%%%%%%%%%%%%%%%%%%%%%%%%%%%%%%%%%%%%%%%%%%%%%%%%%%%%%%%%%%%%%
\begin{figure*}
\centerline{\includegraphics[scale=0.625]{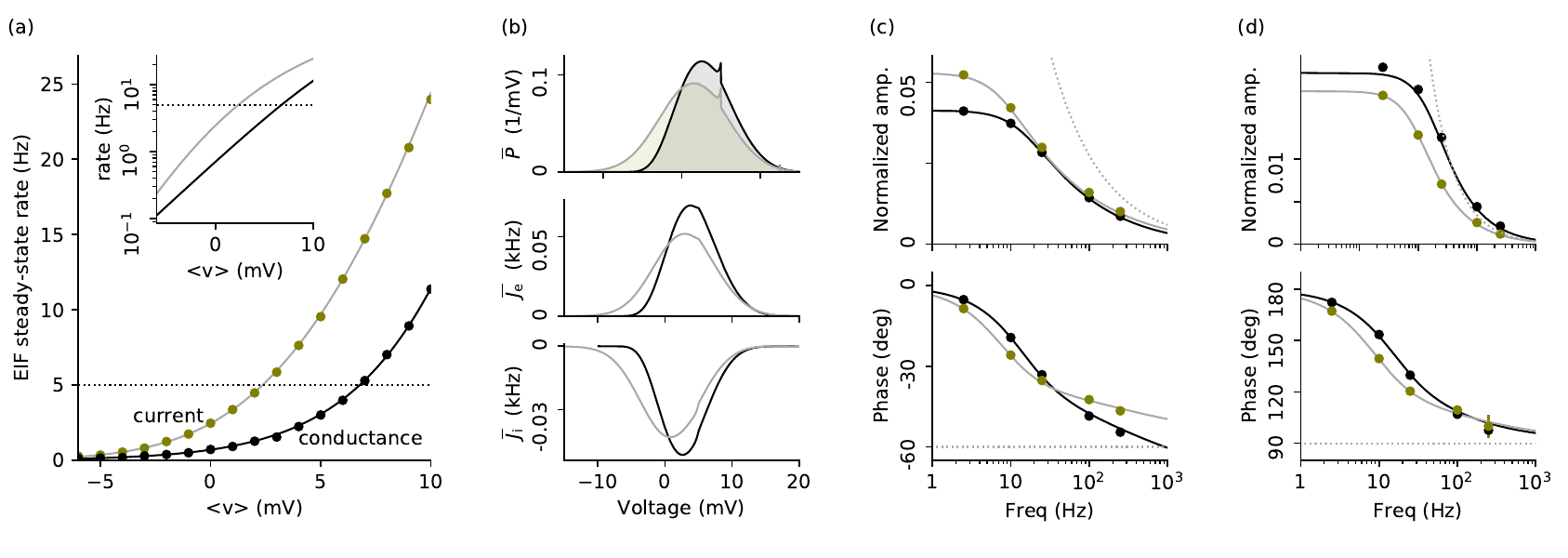}}
\caption{Exponential I\&F neuron driven by conductance or current-based shot noise with excitatory and inhibitory reversal potentials. (a) Steady-state outgoing firing rate with the incoming presynaptic rates $(\SRe,~\SRi)$ the same for the conductance and current-based cases. The presynaptic rates were parameterised, for convenience, from the current-based equations (see Eq. \ref{Curmusig}) for the (subthreshold) voltage mean $\Ex{v}$ with the standard deviation fixed at $\sigma\!=\!5$mV. As for the LIF, again note the near-exponential rise of $\Sr$ for the conductance case seen in the inset. (b) Probability density together with excitatory and inhibitory fluxes for neurons firing at $5$Hz (intersections with dotted line in panel a). The presynaptic rates $(\SRe,~\SRi)$ were $(0.446,0.440)$kHz and $(0.397,0.636)$kHz for conductance and current-based cases, respectively. Note the relatively curtailed distribution near the inhibitory reversal potential $\epsi=-10$mV for the conductance-based case. (c) Amplitude and phase of the firing-rate response to modulation of the presynaptic excitatory rate with asymptotics shown. (d) Same, but for the case of modulation of the inhibitory presynaptic rate. For panels (c-d) the presynaptic rates from panel (b) were used (corresponding to $\Sr\!=\!5$Hz) with all other parameters for the EIF provided in caption to Fig. 1. Simulational results are provided by symbols in panels (a,c,d) and histograms in upper panel (b). Code used to generate the figure is provided in the Supplemental Material \cite{MySupp}.}
\label{fig3}
\end{figure*}

%%%%%%%%%%%%%%%%%%%%%%%%%%%%%%%%%%%%%%%%%%%%%%%%%%%%%%%%%%%%%%%%%%%
%%%%%%%%%%%%%%%%%%%%%%%%%%%%%%%%%%%%%%%%%%%%%%%%%%%%%%%%%%%%%%%%%%%
%%%%%%%%%%%%%%%%%%%%%%%%%%%%%%%%%%%%%%%%%%%%%%%%%%%%%%%%%%%%%%%%%%%
%%%%%%%%%%%%%%%%%%%%%%%%%%%%%%%%%%%%%%%%%%%%%%%%%%%%%%%%%%%%%%%%%%%
% EXPONENTIAL INTEGRATE-AND-FIRE MODEL
%%%%%%%%%%%%%%%%%%%%%%%%%%%%%%%%%%%%%%%%%%%%%%%%%%%%%%%%%%%%%%%%%%% 
%%%%%%%%%%%%%%%%%%%%%%%%%%%%%%%%%%%%%%%%%%%%%%%%%%%%%%%%%%%%%%%%%%%
%%%%%%%%%%%%%%%%%%%%%%%%%%%%%%%%%%%%%%%%%%%%%%%%%%%%%%%%%%%%%%%%%%%
%%%%%%%%%%%%%%%%%%%%%%%%%%%%%%%%%%%%%%%%%%%%%%%%%%%%%%%%%%%%%%%%%%%
\section{\hspace*{-3mm}Exponential Integrate \& Fire model}
The EIF current-voltage term $f(v)$ given by Eq. (\ref{FEIF}) has two zeros determined by the parameters $\vT$ and $\dT$: a stable fixed $\vs$ just above $v\!=\!0$, similar to the LIF model, and an unstable fixed point $\vu$ above $\vT$. These fixed points can be found in terms of Lambert $W$ functions \cite{Toubul2008} but to leading order approximation are roughly
\be
\hspace*{-6mm}\vs\simeq\dT e^{-\vT/\dT} &\mbox{ and }&
\vu\simeq \vT+\dT\log\l(\vT/\dT\r).
\ee
In the absence of synaptic drive, a starting voltage above $\vu$ will take the voltage to infinity in finite time which can be considered the ultimate threshold. However, physiologically, the range of validity of the exponential term is up to $\sim\!10$mV above the unstable fixed point (see inset to Fig. 2a in reference \cite{Badel2008a}) and so it is reasonable, as well as convenient numerically, to set a finite threshold at some value $\vth$ that is above $\vu$ but below $\epse$ and at which point a spike is registered and the voltage is reset to $\vre$. For the EIF model, therefore, the ultimate threshold $\vth$ is set at a higher value than for the LIF model.

%%%%%%%%%%%%%%%%%%%%%%%%%%%%%%%%%%%%%%%%%%%%%%%%%%%%%%%%%%%%%%%%%%%
% Boundary conditions
%%%%%%%%%%%%%%%%%%%%%%%%%%%%%%%%%%%%%%%%%%%%%%%%%%%%%%%%%%%%%%%%%%%
\bigskip\ni\textit{Boundary conditions.} At threshold the firing rate is given by the flux $J(\vth,t)$. However, unlike for the LIF model, in this voltage range both the (positive) deterministic component of the flux and the excitatory flux contribute to crossing the threshold
\be
\hspace*{-5mm} r(t)=J(\vth,t)&=&f(\vth)P(\vth,t)+\Je(\vth,t) \\
\hspace*{-5mm} \mbox{ with }\Ji(\vth,t)&=&0
\ee
where the second result is due to there being no neuronal trajectories above $\vth$ due to the instantaneous reset condition. Unlike the LIF model, there is no equality between the total flux and excitatory flux at threshold so the probability density at threshold does not vanish. Additionally, the flux equation (\ref{flux}) reduces to $J\!=\!\Je\!+\!\Ji$ at the two fixed points (when $f(v)\!=\!0$).

%%%%%%%%%%%%%%%%%%%%%%%%%%%%%%%%%%%%%%%%%%%%%%%%%%%%%%%%%%%%%%%%%%%
%%%%%%%%%%%%%%%%%%%%%%%%%%%%%%%%%%%%%%%%%%%%%%%%%%%%%%%%%%%%%%%%%%%
% Steady-state for the EIF with current-based case
%%%%%%%%%%%%%%%%%%%%%%%%%%%%%%%%%%%%%%%%%%%%%%%%%%%%%%%%%%%%%%%%%%%
%%%%%%%%%%%%%%%%%%%%%%%%%%%%%%%%%%%%%%%%%%%%%%%%%%%%%%%%%%%%%%%%%%%
\subsection*{EIF steady-state rate}\vspace*{-2mm}
Analytical solutions to the steady-state master equation (Eqs. \ref{dJdv0con}-\ref{dJidv0con}) driven by excitatory and inhibitory shot noise were not forthcoming and so again a numerical scheme was developed. However, some care is needed due to the presence of the fixed points at $\vs$ and $\vu$ where $f(v)$ vanishes. This requires that the steady-state master equation is integrated in three domains
\be
[\epsi \xrightarrow{(i)}\vs]~[\vs \xleftarrow{(ii)} \vu]~ [\vu \xrightarrow{(iii)} \vth]\label{EIFdomains}
\ee
with matching across the domain interfaces $\vs$ and $\vu$ as well as the boundary conditions at $\epsi$ and $\vth$ respected. It is convenient to first solve in domain (iii) from the unstable fixed point up to the threshold, then in domain (ii) from the unstable fixed point down to the stable fixed point and finally in domain (i) from the inhibitory reversal potential to the stable fixed point. Like for the LIF model, the only inhomogeneous term is in Eq. (\ref{dJdv0con}) and is proportional to $\Sr$. All downstream quantities will therefore be proportional to this unknown quantity and it can be scaled out (lower-case letters are used for these quantites). The scaled flux is again $\Sj\!=\!\theta(v\!-\!\vre)$. It remains only to solve for the two synaptic fluxes $(\Sje,\Sji)$ with the probability density written in terms of these quantities using the flux equation in the steady state, as was done for the LIF. 

Numerical stability requires that integration of the equation set is in the direction away from the unstable fixed point and towards the stable fixed point. This adds the complication that the ratio of excitation to inhibition is unknown at $\vu$. As stated above, the method is to first solve the equations in region (iii) and to do so twice. For the first solution $(\Sje^A,\Sji^A)$ initial conditions at $\vu$ of $(1,0)$ are chosen using $\Sp^A=(1-\Sje^A-\Sji^A)/f$. For the second solution $(\Sje^B,\Sji^B)$ the initial conditions at $\vu$ are $(1,-1)$ and $\Sp^B=-(\Sje^B+\Sji^A)/f$ is used. The requirement of a vanishing inhibitory flux at threshold is then used to get the correct mix of these solutions 
\be
\Sji^A(\vth)+\kappa\Sji^B(\vth)=\Sji(\vth)=0
\ee 
fixing the constant $\kappa$. Once achieved, the correct combination of $A$ and $B$ solutions can be integrated in region $(ii)$ from $\vu$ down to the stable fixed point $\vs$. Matching at $\vs$ is achieved by rescaling the solution from region $(i)$, which is found in the same way as for the LIF model. Finally, the steady-state rate $\Sr$ is recovered from the normalisation condition on $\SP=\Sr\Sp$. Examples of the steady-state rate (Fig. \ref{fig3}a), density and flux distributions (Fig. \ref{fig3}b) are provided for the conductance-based case with comparison to the case of current-based additive shot noise.  

%%%%%%%%%%%%%%%%%%%%%%%%%%%%%%%%%%%%%%%%%%%%%%%%%%%%%%%%%%%%%%%%%%%
%%%%%%%%%%%%%%%%%%%%%%%%%%%%%%%%%%%%%%%%%%%%%%%%%%%%%%%%%%%%%%%%%%%
% Firing-rate response for the EIF
%%%%%%%%%%%%%%%%%%%%%%%%%%%%%%%%%%%%%%%%%%%%%%%%%%%%%%%%%%%%%%%%%%%
%%%%%%%%%%%%%%%%%%%%%%%%%%%%%%%%%%%%%%%%%%%%%%%%%%%%%%%%%%%%%%%%%%%
\subsection*{EIF firing-rate response}\vspace*{-2mm}
The master equation at the level of modulation, given by Eqs. (\ref{dJdv1con}-\ref{dJidv1con}) again appears difficult to solve analytically. However, a numerical approach similar to that used for the steady-state rate can be developed. As for the LIF case, at the level of modulation, the numerical solution is relatively involved to construct due to the multiple inhomogeneous terms and difficulty in accounting for boundary conditions and the fixed points of $f(v)$ for the EIF model: the method is therefore described in detail in the Appendix. Example results for the numerical solutions are provided in Fig. \ref{fig3}c, \ref{fig3}d and compared to the matched current-based shot-noise model.

%%%%%%%%%%%%%%%%%%%%%%%%%%%%%%%%%%%%%%%%%%%%%%%%%%%%%%%%%%%%%%%%%%%
% Modulation for the EIF
%%%%%%%%%%%%%%%%%%%%%%%%%%%%%%%%%%%%%%%%%%%%%%%%%%%%%%%%%%%%%%%%%%%
\bigskip\ni\textit{Modulation at high frequencies.} For the case of excitatory current-based shot noise, an analytical approach was developed that allowed the rate-response at high frequencies to be related to steady-state synaptic fluxes \cite{Richardson2018}. A similar framework can be developed for the case of combined excitatory and inhibitory flux and is now presented. 

First, the notion of the escape time $\EscT$ is introduced. In the absence of the synaptic input term $s(v,t)$ in Eq. (\ref{dvdt}), the time taken for the voltage to diverge to infinity from a voltage above the unstable fixed point $v\!>\!\vu$ is
\be
\EscT=\int_v^\infty\frac{dv}{f(v)} &\mbox{ where } & \frac{d\EscT}{dv}=-\frac{1}{f}. \label{Tesc}
\ee 
Here $v\!=\!\vu$ corresponds to $\EscT\!=\!\infty$ and $v\!=\!\infty$ corresponds to $\EscT\!=\!0$. Separately, in the same domain of $v\!>\!\vu$, the modulated continuity equation (Eq. \ref{dJdv1con}) can be rewritten for $\MP$ and substituted into the modulated flux equation $\MJ\!=\!f\MP+\MJe+\MJi$ to give
\be
\frac{f}{i\w}\frac{d\MJ}{dv}+\MJ=\MJe+\MJi.
\ee 
These two results can now be combined by using the derivative form in Eq. (\ref{Tesc}) to convert the differential equation in $v$ to one in $\EscT(v)$. Making use of the boundary conditions that $\MJ\!\to\!\Mr$ in the limit $v\!\to\!\infty$, $\EscT\!\to\!0$ and $\MJ\!\to\!\MJe+\MJi$ at $v\!=\!\vu$, $\EscT\!\to\!\infty$ gives the result
\be
\Mr=\int_0^\infty d\EscT e^{-i\w \EscT}\l(\frac{d\MJe}{d\EscT}+\frac{d\MJi}{d\EscT}\r). \label{Ttransform}
\ee
At high frequencies, using Eq. (\ref{JeJiAsym}), the asymptotics can be written in terms of the steady state fluxes 
\be
\Mr\simeq\int_0^\infty d\EscT e^{-i\w \EscT}\l(\frac{d\SJe}{d\EscT}+\frac{d\SJi}{d\EscT}\r) \label{TtransformSS}
\ee
allowing for the high-frequency asymptotics - a parameterisation of the dynamics - to be derived if an analytically convenient form for the steady-state synaptic fluxes can
be found in terms of the escape time $\EscT(v)$. This is in the nature of a fluctuation-dissipation relation (though see reference \cite{Lindner2022} for a complete relation for neuronal systems). Unfortunately, this does not appear to be as straightforward as for the current-based case \cite{Richardson2018} in which the asymptotics had a dependency on the relation between the ratio of the mean excitatory amplitude and the spike sharpness $\dT$. However, the broad behaviour remains as can be seen in Fig. \ref{fig3}c, \ref{fig3}d in which the rate-response of the models with conductance or and current-based drive have similar forms over moderate frequencies.

%%%%%%%%%%%%%%%%%%%%%%%%%%%%%%%%%%%%%%%%%%%%%%%%%%%%%%%%%%%%%%%%%%%
%%%%%%%%%%%%%%%%%%%%%%%%%%%%%%%%%%%%%%%%%%%%%%%%%%%%%%%%%%%%%%%%%%%
%%%%%%%%%%%%%%%%%%%%%%%%%%%%%%%%%%%%%%%%%%%%%%%%%%%%%%%%%%%%%%%%%%%
%%%%%%%%%%%%%%%%%%%%%%%%%%%%%%%%%%%%%%%%%%%%%%%%%%%%%%%%%%%%%%%%%%%
% Discussion
%%%%%%%%%%%%%%%%%%%%%%%%%%%%%%%%%%%%%%%%%%%%%%%%%%%%%%%%%%%%%%%%%%% 
%%%%%%%%%%%%%%%%%%%%%%%%%%%%%%%%%%%%%%%%%%%%%%%%%%%%%%%%%%%%%%%%%%%
%%%%%%%%%%%%%%%%%%%%%%%%%%%%%%%%%%%%%%%%%%%%%%%%%%%%%%%%%%%%%%%%%%%
%%%%%%%%%%%%%%%%%%%%%%%%%%%%%%%%%%%%%%%%%%%%%%%%%%%%%%%%%%%%%%%%%%%
\section{Discussion} \vspace*{-2mm}
The population-level framework for linear and non-linear neurons driven by exponentially distributed excitatory and inhibitory synaptic condutances was examined. For this biophysically reasonable choice of conductance distribution it was demonstrated that the generic integro-differential master equation (see reference \cite{Nykamp2000}) can be reduced to a set of more tractable differential equations. This description of the master equation allowed for the development of an efficient scheme for the numerical derivation of the steady-state density distributions and rate as well as those at the level of the firing-rate response to weakly modulated presynaptic rates. The analyses of these equations presented here extend treatments of the effects of synaptic conductance on the LIF and EIF firing rates previously derived in the small-amplitude appproximation \cite{Johannesma1968,Richardson2004,Richardson2007}. They also extend previous results for finite-amplitude synaptic drive for the LIF \cite{Richardson2010} and EIF \cite{Richardson2018} models where synapses were treated as current-based and the noise was therefore additive rather than conductance-based and multiplicative. 

For the case of additive shot-noise, it was possible to derive analytical forms for the steady-state and modulated rates \cite{Richardson2010} for the LIF with excitatory and inhibitory input. These results were later extended to non-exponentially distributed inhibition \cite{Angulo-Garcia2017,Olmi2017}. Unfortunately, that analytical approach, which used bilateral Laplace transforms, does not appear to extend straightforwardly to the case of reversal potentials: analytical solutions for these quantities remain a topic for future research. Interestingly, the steady-state firing rate as shown in the inset to Fig. \ref{fig2}a appears close to exponential, suggesting that there may nevertheless be a simple approximation, if not a full solution, to the steady-state rate with combined excitation and inhibition

An analytical solution for the rate-response for the LIF model was similarly illusive, though the high-frequency asymptotics could be obtained. The difficulty in finding an analytical solution of course also extends to the more detailed EIF model, though again an intriguing near-exponential rise in the steady-state firing rate can be seen in the inset to Fig. \ref{fig3}a This suggests that a simple solution (or, at the least, an accurate approximation) to the steady-state rate might be possible.

It was previously shown \cite{Richardson2018} that, at the level of the firing-rate response for the EIF model with current-based drive, there is an interplay between the mean amplitude $\ame$ and the sharpness of the spike $\dT$ in setting the asymptotic high-frequency exponent. Though it was again possible to show that the asymptotics of the high-frequency response can be related to the steady-state excitatory flux - a form of fluctuation-dissipation relation - the simplification of the excitatory flux for the case of reversal potentials was not so straightforward to derive as for the current-based case because the mean synaptic amplitude is conditional on voltage. However, the dependency on the rapidity of the response as a ratio of the amplitude around threshold and the spike sharpness broadly holds over moderate frequencies (Fig. \ref{fig3}c).

These results represent an initial foray into models of linear and non-linear integrate-and-fire models with exponentially distributed conductances. While it is clear from the structure of the master equation that the steady-state voltage distribution can be trivially derived in integral form for both the LIF and EIF in the case of excitation only, in the context of synaptic reversal potentials it is the inclusion of inhibition that is of relevance. The solution for the steady-state rate for combined excitation and inhibition as well as the firing-rate response remain therefore an important theoretical goal. 

A further case of interest would be to relate a changing amplitude distribution to the role of short-term synaptic plasticity \cite{Abbott1997,Tsodyks1997}. This has been analysed in the Gaussian approximation \cite{Rosenbaum2012,Bird2014} but would be interesting to examine as a driver for how the amplitude distribution changes as a function of presynaptic activity, the history dependence of inputs at particular fibres and the effect of non-Poissonain statistics on the population equations \cite{Ly2009,Lai2017}. The high-frequency asymptotics of the rate response  are sensitive to the interplay of the spike-sharpness and mean synaptic amplitude \cite{Richardson2018}. Including a mechanism like short-term plasticity provides a motivation for analysing how the rapidity of the neuronal response can be modulated by the level of presynaptic network activity and is a physiologically valid question for future research.

\section*{APPENDIX A: NUMERICS FOR THE\\[1mm] FIRING-RATE RESPONSE} \vspace*{-2mm}
The numerical solution for the steady-state master equation (\ref{dJdv0con}-\ref{dJidv0con}) for the conductance-based shot-noise LIF and EIF models were described in the main text. The approach is an extension of the Threshold Integration methods developed for Gaussian-white noise \cite{Richardson2007}, LIF with additive shot noise \cite{Richardson2010} and EIF with additive excitatory shot noise \cite{Richardson2018}. In this section, the approach for modulated excitatory or inhibitory conductance-based shot noise is outlined. The general components common to the LIF and EIF are first described and then the boundary conditions specific to the two models explained in detail. All Julia code \cite{Bezanson2017} is provided in the Supplemental Material \cite{MySupp}.

The method now presented is similar to that used for the steady state, though in the modulation case there are three inhomogeneous components proportional to $\Mr$, $\MRe$ and $\MRi$ in the region (ii) for the LIF and (ii) and (iii) for the EIF (see Eqs. \ref{LIFdomains} and \ref{EIFdomains}). The system is linear so it suffices to solve the problem with either excitatory modulation or inhibitory modulation separately, with effects of combinations of the remaining two modulations simply adding at the population level in the weak modulation approximation. Throughout the following, the example of excitatory modulation will be used without loss of generality. There are therefore two inhomogeneous terms that are considered together: that of $\Mr$, $\MRe$ allowing the solutions for the various fluxes to be separated into a sum of two subsolutions, like for the total flux:
\be
\MJ=\Mr\Mj^r+\MRe\Mj^\me.
\ee
Equations (\ref{dJdv1con}-\ref{dJidv1con}) can be resolved into two sets of equations with variables $(\Mj^r,\Mje^r,\Mji^r)$ and $(\Mj^\me,\Mje^\me,\Mji^\me)$. The first set has $\Mr\!=\!1$ and $\MRe\!=\!\MRi\!=\!0$ and the second set has $\Mr\!=\!0$ and $\MRe\!=1$ with $\MRi\!=\!0$ still because here only excitatory modulation is considered. The solution approach is relatively straightforward for the LIF in region (ii) because the boundary conditions can be imposed at the threshold, which is the starting point for the integration back to the stable fixed point. For the EIF, however, there is an added complication because it is necessary to integrate from the unstable fixed point (where boundary conditions are not directly specified) up to the threshold where they are - the strategy for handling this complication is explained in detail later.

In domain (i), from the reversal potential to the stable fixed point for the LIF or EIF, it is numerically convenient to run the integration from a lower-bound $\vlb$ that is slightly above the inhibitory reversal potential $\epsi$. A zero-flux condition is imposed at $\vlb$. The excitatory flux is zero here (as there are no neuronal trajectories below) so the inhibitory flux balances the deterministic component $\Ji(\vlb)=-fP$. Calling the magnitude of this inhibitory flux $\Mq$, the equations in region (i) for both the LIF and EIF can be resolved into two components. So, again using the total flux as an example: 
\be
\MJ=\Mq\Mj^q+\MRe\Mj^\me. \label{qbit}
\ee
Once each subsolution has been obtained with the appropriate boundary conditions at $\vlb$ (see later) the equation in domain (i) can be scaled to match with that in (ii) at the stable fixed point ($v=0$ for the LIF, $v=\vs$ for the EIF) 
\be
(\Mq\Mje^q+\MRe\Mje^\me)|_{-}=(\Mr\Mje^r+\MRe\Mje^\me)|_{+} \label{Bounde}\\
(\Mq\Mji^q+\MRe\Mji^\me)|_{-}=(\Mr\Mji^r+\MRe\Mji^\me)|_{+} \label{Boundi}
\ee
where the $\pm$ means either just below or above the stable fixed point. These two equations can be trivially solved to finally provide $\Mq$ and the desired modulatory rate $\Mr$ in response to presynaptic excitatory modulation. The case of inhibitory modulation is derived using an identical approach but with $\MRe\!=\!0$ and $\MRi\!=\!1$. The specific boundary conditions for the LIF or EIF model are now described in more detail.

%%%%%%%%%%%%%%%%%%%%%%%%%%%%%%%%%%%%%%%%%%%%%%%%%%%%%%%%%%%%%%%%%%%
%%%%%%%%%%%%%%%%%%%%%%%%%%%%%%%%%%%%%%%%%%%%%%%%%%%%%%%%%%%%%%%%%%%
% Appendix: numerical rate-response for the LIF
%%%%%%%%%%%%%%%%%%%%%%%%%%%%%%%%%%%%%%%%%%%%%%%%%%%%%%%%%%%%%%%%%%%
%%%%%%%%%%%%%%%%%%%%%%%%%%%%%%%%%%%%%%%%%%%%%%%%%%%%%%%%%%%%%%%%%%%
\subsection*{LIF boundary conditions}
For the LIF there are two domains to integrate over (see Eq. \ref{LIFdomains}). Starting with domain (ii), first integrate $0\leftarrow\vth$ with initial conditions $(1,1,0)$ for the $(\Mj^r,\Mje^r,\Mji^r)$ solution and $(0,0,0)$ for the $(\Mj^\me,\Mje^\me,\Mji^\me)$ solution. Then integrate in domain (i) up from $\vlb\rightarrow0$ with initial conditions $(0,0,-1)$ for the $(\Mj^q,\Mje^q,\Mji^q)$ solution and $(0,0,0)$ for the $(\Mj^\me,\Mje^\me,\Mji^\me)$ solution. The matching criterion then gives $\Mr$ as required from the solution of Eqs. (\ref{Bounde}, \ref{Boundi}). The approach for inhibitory modulation is analagous.

%%%%%%%%%%%%%%%%%%%%%%%%%%%%%%%%%%%%%%%%%%%%%%%%%%%%%%%%%%%%%%%%%%%
%%%%%%%%%%%%%%%%%%%%%%%%%%%%%%%%%%%%%%%%%%%%%%%%%%%%%%%%%%%%%%%%%%%
% Appendix: numerical rate-response for the EIF
%%%%%%%%%%%%%%%%%%%%%%%%%%%%%%%%%%%%%%%%%%%%%%%%%%%%%%%%%%%%%%%%%%%
%%%%%%%%%%%%%%%%%%%%%%%%%%%%%%%%%%%%%%%%%%%%%%%%%%%%%%%%%%%%%%%%%%%
\subsection*{EIF boundary conditions}
The method is similar to that described above for the LIF except that, like for the steady-state case, the solution in the domain (iii) from $\vu\to\vth$ needs to be properly constructed before the solutions in the other domains (ii) and (i) are found. Again, for simplicity of exposition, a case is considered where there is excitatory modulation only.

%%%%%%%%%%%%%%%%%%%%%%%%%%%%%%%%%%%%%%%%%%%%%%%%%%%%%%%%%%%%%%%%%%%
% Appendix: EIF modulation regeion (iii)
%%%%%%%%%%%%%%%%%%%%%%%%%%%%%%%%%%%%%%%%%%%%%%%%%%%%%%%%%%%%%%%%%%%
\bigskip\ni\textit{Region (iii).} Stability requires integration to be in the direction $\vu\to\vth$. However, the boundary conditions, that $\MJ\!=\!\Mr$ and $\MJi\!=\!0$, are imposed at $\vth$. Hence, linearly independent solutions with different initial conditions at $\vu$ need to be appropriately combined so that the desired conditions at $\vth$ are met. First, a solution $(\Mj^\mE,\Mje^\mE,\Mji^\mE)$ to Eqs. (\ref{dJdv1con}-\ref{dJidv1con}) with $\MRe\!=\!1$, $\MRi\!=\!0$ are integrated from $\vu\to\vth$ with initial conditions $(0,0,0)$. Two other solutions, $(\Mj^A,\Mje^A,\Mji^A)$ and $(\Mj^B,\Mje^B,\Mji^B)$ are then integrated with $\MRe,\MRi\!=\!0,0$, with initial conditions $(1,1,0)$ and $(0,1,-1)$ respectively. The solution that deals with the inhomogeneous term proportional to $\MRe$ can now be constructed as a linear combination of these three solutions such that, for example, the total flux is written
\be
\Mj^\me=a\Mj^A+b\Mj^B+\Mj^\mE.
\ee 
Given freedom to vary $a$ and $b$ (the two homogeneous solutions), a scenario is chosen where both the total and inhibitory flux vanish at threshold for this component
\be
0&=&(a\Mj^A+b\Mj^B+\Mj^\me)|_\vth ~\mbox{ and} \\
0&=&(a\Mji^A+b\Mji^B+\Mji^\me)|_\vth. 
\ee
This fixes the values of $a$ and $b$ in terms of the integrated solutions at $\vth$. The $\Mr$ component $(\Mj^r,\Mje^r,\Mji^r)$ can also be constructed from the $A$ and $B$ subsolutions, so for example for the total flux 
\be
\Mj^r=c\Mj^A+d\Mj^B.
\ee
Given the way the $\MRe$ component was constructed to have vanishing total flux at threshold, the $\Mr$ component now has to satisfy the boundary conditions $\Mj=1$ as well as $\Mji=0$. Hence, the requirements 
\be
1&=&(c\Mj^A+d\Mj^B)|_\vth ~\mbox{ and }\\
0&=&(c\Mji^A+d\Mji^B)|_\vth 
\ee
that together fix $c$ and $d$. These provide the correct combination of solutions for the two inhomogeneous solutions proportional to $\MRe$ and $\Mr$ in domain (iii).

%%%%%%%%%%%%%%%%%%%%%%%%%%%%%%%%%%%%%%%%%%%%%%%%%%%%%%%%%%%%%%%%%%%
% Appendix: EIF modulation regeion (ii)
%%%%%%%%%%%%%%%%%%%%%%%%%%%%%%%%%%%%%%%%%%%%%%%%%%%%%%%%%%%%%%%%%%%
\bigskip\ni\textit{Region (ii).} Here the integration is downwards $(\vs\!\leftarrow\!\vu)$ with the initial conditions from the previous case being
\be
(\Mj^\me,\Mje^\me,\Mji^\me)|_{\vu}&=&a(1,1,0)+b(0,1,-1) \nonumber \\
(\Mj^r,\Mje^r,\Mji^r)|_{\vu}&=&c(1,1,0)+d(0,1,-1).
\ee
It should be remembered that the initial conditions for the solution $(\Mj^\mE,\Mje^\mE,\Mji^\mE)$ were $(0,0,0)$ and so don't contribute to the first of the conditions above. Both sets of equations, for the $\Mr$ and $\MRe$ inhomogeneous solutions, are integrated down to $\vs$. Note that because $\vre$ is in this domain, the integration for the $\Mr$ solution includes the Dirac-delta function in Eq. \ref{dJdv1con}.

%%%%%%%%%%%%%%%%%%%%%%%%%%%%%%%%%%%%%%%%%%%%%%%%%%%%%%%%%%%%%%%%%%%
% Appendix: EIF modulation region (i)
%%%%%%%%%%%%%%%%%%%%%%%%%%%%%%%%%%%%%%%%%%%%%%%%%%%%%%%%%%%%%%%%%%%
\ni\textit{Region (i).} As explained above, a lower bound $\vlb$ is imposed just above the inhibitory flux. At this point, all fluxes will be zero except for the inhibitory one which is set as being proportional to a quantity $\Mq$ (see Eq. \ref{qbit}). The initial conditions for the solution $(\Mj^q,\Mje^q)$ are $(0,0,-1)$ and $(0,0,0)$ for $(\Mj^\me,\Mje^\me,\Mji^\me)$. Finally, $\Mr$ and $\Mq$ are determined from the linear equations (\ref{Bounde},\ref{Boundi}) given earlier thereby completing the numerical solution.

%%%%%%%%%%%%%%%%%%%%%%%%%%%%%%%%%%%%%%%%%%%%%%%%%%%%%%%%%%%%%%%%%%%
%%%%%%%%%%%%%%%%%%%%%%%%%%%%%%%%%%%%%%%%%%%%%%%%%%%%%%%%%%%%%%%%%%%
%%%%%%%%%%%%%%%%%%%%%%%%%%%%%%%%%%%%%%%%%%%%%%%%%%%%%%%%%%%%%%%%%%%
%%%%%%%%%%%%%%%%%%%%%%%%%%%%%%%%%%%%%%%%%%%%%%%%%%%%%%%%%%%%%%%%%%%
% Appendix B: current-based synapses
%%%%%%%%%%%%%%%%%%%%%%%%%%%%%%%%%%%%%%%%%%%%%%%%%%%%%%%%%%%%%%%%%%% 
%%%%%%%%%%%%%%%%%%%%%%%%%%%%%%%%%%%%%%%%%%%%%%%%%%%%%%%%%%%%%%%%%%%
%%%%%%%%%%%%%%%%%%%%%%%%%%%%%%%%%%%%%%%%%%%%%%%%%%%%%%%%%%%%%%%%%%%
%%%%%%%%%%%%%%%%%%%%%%%%%%%%%%%%%%%%%%%%%%%%%%%%%%%%%%%%%%%%%%%%%%%
\section*{APPENDIX B: ADDITIVE SHOT NOISE} \vspace*{-2mm}
The framework for additive, current-based shot noise with exponentially distributed amplitudes was developed in reference \cite{Richardson2010} for the LIF driven by excitatory and inhibitory input and later for the EIF driven by excitatory input only \cite{Richardson2018}. In this section, the analysis in \cite{Richardson2018} is extended to combined excitation and inhibition. For current-based drive, the amplitude distribution is independent of voltage with the drive term in
Eq. (\ref{dvdt}) written as
\be
S(t)=\sum_{k}a^\me_{k}\delta(t-t^\me_{k})+\sum_{k}a^\mi_{k}\delta(t-t^\mi_{k}) \label{Scurr}
\ee
where $a^\me_{k}$ is the amplitude and $t^\me_{k}$ the time of the $k$th
excitatory input, with a similar definition for inhibition. The voltage amplitudes are drawn from exponential distributions $A_\me(a)$ with mean $a_\me\!>\!0$, using excitation as an example, so that 
\be
\hspace*{-3mm} A_\me(a)=\theta(a)\frac{e^{-a/a_\me}}{a_\me} ~&
\mbox{and} &~ T_\me(a)=\theta(a) e^{-a/a_\me}. \label{Aexp}
\ee
The tail distribution $T_\me(a)$ is the probability that an
amplitude is greater than $a$ and, unlike for the conductance-based case, is independent of voltage.  The inhibitory jumps are also exponential distributed, though with $a_\mi<0$. Example distributions
are provided in Fig. \ref{fig1}b lower panel for comparison to the case with reversal potentials. 

%%%%%%%%%%%%%%%%%%%%%%%%%%%%%%%%%%%%%%%%%%%%%%%%%%%%%%%%%%%%%%%%%%%
%%%%%%%%%%%%%%%%%%%%%%%%%%%%%%%%%%%%%%%%%%%%%%%%%%%%%%%%%%%%%%%%%%%
% Appendix: Flux and continuity equations
%%%%%%%%%%%%%%%%%%%%%%%%%%%%%%%%%%%%%%%%%%%%%%%%%%%%%%%%%%%%%%%%%%%
%%%%%%%%%%%%%%%%%%%%%%%%%%%%%%%%%%%%%%%%%%%%%%%%%%%%%%%%%%%%%%%%%%%
\subsection*{Synaptic flux equations for additive shot noise}\vspace*{-2mm}
The excitatory flux across a voltage $v$ is equal to the rate
that jumps from all values of the voltage $w<v$ cross it. For current-based shot noise, this is just a convolution of the probability density and tail distribution, which for exponentially distributed input can be written
\be
\Je(v,t)=\Re\int_{-\infty}^vdwP(w)e^{-(v-w)/\ame} \label{JeI}
\ee
with a similar form derivable for inhibition. This integral form can be recognized as the solution of a first-order linear differential equation for $\Je$ with the density $P$ acting as an inhomogeneous term, with a similar result for inhibition
\be
\frac{\p J_\me}{\p v}+\frac{J_\me}{a_e}=R_\me P &\mbox{ and }& \frac{d
  J_\mi}{dv}+\frac{J_\mi}{a_\mi}=R_\mi P \label{IJeJi}.
\ee
Together with the continuity and flux equation (Eqs. \ref{cont}-\ref{flux}) these differential equations constitute the master equation that fully describes the dynamics of an ensemble of neurons subject to current-based shot noise with exponentially distributed amplitudes. 

%%%%%%%%%%%%%%%%%%%%%%%%%%%%%%%%%%%%%%%%%%%%%%%%%%%%%%%%%%%%%%%%%%%
%%%%%%%%%%%%%%%%%%%%%%%%%%%%%%%%%%%%%%%%%%%%%%%%%%%%%%%%%%%%%%%%%%%
% Appendix: LIF for current-based synapses
%%%%%%%%%%%%%%%%%%%%%%%%%%%%%%%%%%%%%%%%%%%%%%%%%%%%%%%%%%%%%%%%%%%
%%%%%%%%%%%%%%%%%%%%%%%%%%%%%%%%%%%%%%%%%%%%%%%%%%%%%%%%%%%%%%%%%%%
\subsection*{LIF with additive shot noise} \vspace*{-2mm}
The master equation for this model can be solved to find both the steady-state rate and firing-rate response through a bilateral Laplace transform of the voltage \cite{Richardson2010}. Both rates can be written in terms of the voltage moment-generating function $Z(s)$ 
\be
\frac{1}{Z(s)}=\l(1\!-\!a_\me s\r)^{\tau R_\me} \l(1\!-\!a_\mi s\r)^{\tau
  R_\mi}
\ee
giving the steady-state voltage mean and variance as
\be
\hspace*{-7mm}\Ex{v}\!=\!\ame\tau\SRe+\ami\tau\SRi &\mbox{and}& \mathrm{Var}(v)\!=\!\ame^2\tau\SRe+\ami^2\tau\SRi. \label{Curmusig}
\ee
Fixing these two quantities specifies $\SRe$ and $\SRi$ which is used as the basis of the steady-state rate in Figs \ref{fig2}a, \ref{fig2}b. The steady-state firing rate can be written in terms of an integral over the generating function
\be
\hspace{-2mm}\frac{1}{\tau r}\!=\!\!\!\int_0^{1/a_e}\!\frac{ds}{s}\frac{1}{Z(s)} \!\l(\frac{e^{sv_{th}}}{1\!-\!a_es}\!-\!e^{s
  v_{re}}\!\r)\!. \label{rateLIFsteady}
\ee
This is the shot-noise generalization of the simplified form \cite{Brunel1999} of the Ricciardi formula \cite{Ricciardi1977}.  The corresponding firing-rate response for weak modulation of either the excitatory or inhibitory presynaptic rate is written 
\be
\hspace*{-5mm}\Mr_\kappa=\MR_\kappa\tau r\frac{\int_0^{1/a_e}\frac{ds}{s} \frac{1}{Z(s)} 
\l(\frac{e^{sv_{th}}}{1\!-\!a_es}\!-\!e^{s
  v_{re}}\r)\int_0^s\frac{dc a_\kappa c^{i\w\tau}}{1-a_\kappa c}  }{\int_0^{1/a_e}\frac{ds}{s} \frac{1}{Z(s)} 
\l(\frac{e^{sv_{th}}}{1\!-\!a_es}\!-\!e^{s
  v_{re}}\r)e^{i\w t}} \label{LIF1sol}
\ee
where $\kappa=\me,\mi$ for excitation or inhibition. Though these analytical forms exist for the LIF with additive shot noise, it is often more convenient to generate the solutions numerically using a similar method to that described in Appendix A for conductance-based shot noise. The principal difference is that the lower bound is not constrained to be above any reversal potential for inhibition but rather chosen to be sufficiently low that it has little effect on the results. Finally, in the limit of high frequencies, the asymptotics can be shown to be 
\be
\Mre\simeq \Sr\frac{\MRe}{R_e} & \mbox{ and } & \Mri\simeq
\Sr\frac{\MRi}{i\w}\frac{\ami}{\ame-\ami}
\ee
for excitatory and inhibitory modulation.

%%%%%%%%%%%%%%%%%%%%%%%%%%%%%%%%%%%%%%%%%%%%%%%%%%%%%%%%%%%%%%%%%%%
%%%%%%%%%%%%%%%%%%%%%%%%%%%%%%%%%%%%%%%%%%%%%%%%%%%%%%%%%%%%%%%%%%%
% Appendix: EIF for current-based synapses
%%%%%%%%%%%%%%%%%%%%%%%%%%%%%%%%%%%%%%%%%%%%%%%%%%%%%%%%%%%%%%%%%%% 
%%%%%%%%%%%%%%%%%%%%%%%%%%%%%%%%%%%%%%%%%%%%%%%%%%%%%%%%%%%%%%%%%%%
\subsection*{EIF with additive shot noise} \vspace*{-2mm}
Unfortunately, the Laplace-transform solution used for the LIF does not transfer easily to the EIF with additive shot noise. The solutions are therefore obtained numerically, in the same way as for the conductance-based case but, again, with the lower bound $\vlb$ no-longer constrained by any inhibitory reversal potential $\epsi$. Though a numerical approach is required for the full solution, the firing-rate response asymptotics can nevertheless be obtained, as was shown for the case of the EIF driven by only excitatory current-based shot noise \cite{Richardson2018}. In the following section these results are updated for the case of excitatory and inhibitory input.

%%%%%%%%%%%%%%%%%%%%%%%%%%%%%%%%%%%%%%%%%%%%%%%%%%%%%%%%%%%%%%%%%%%
% Appendix: EIF excite asymptotics
%%%%%%%%%%%%%%%%%%%%%%%%%%%%%%%%%%%%%%%%%%%%%%%%%%%%%%%%%%%%%%%%%%%
\bigskip\ni\textit{Excitatory modulation at high frequencies.} For the excitatory-only drive it was shown \cite{Richardson2018} that the asymptotics for excitatory modulation depend non-trivially on the ratio of the excitatory synaptic amplitude $\ame$ to the spike sharpness $\dT$. The method used in reference \cite{Richardson2018} is identical when background steady-state inhibition is included with the results taking the same form
\be
\Mr\simeq\MRe\frac{r\tau}{i\w\tau}\frac{\ame}{\dT-\ame} &\mbox{
  for $\ame<\dT$}& \\
\Mr\simeq\MRe\frac{r\tau}{i\w\tau}\log(i\w\tau) &\mbox{ for $\ame=\dT$}& \\
\Mr\simeq\MRe\frac{r\tau
  I_e}{(i\w\tau)^{\dT/\ame}}\Gamma\l(\frac{\dT}{\ame}+1\r) &\mbox{
  for $\ame>\dT$.}& \hspace{1cm}
\ee
The integral $I_\me$ is a function of the steady-state density
\be
I_{\me}\!=\!\int_{-\infty}^\infty du e^{(u-v_T)/a_e}\SP(u)/\Sr\tau
\ee
and therefore includes the presence of both excitatory and inhibitory drive. 

%%%%%%%%%%%%%%%%%%%%%%%%%%%%%%%%%%%%%%%%%%%%%%%%%%%%%%%%%%%%%%%%%%
% Appendix: EIF inhib asymptotics
%%%%%%%%%%%%%%%%%%%%%%%%%%%%%%%%%%%%%%%%%%%%%%%%%%%%%%%%%%%%%%%%%%%
\bigskip\ni\textit{Inhibitory modulation at high frequencies.} As inhibitory modulation was not considered in reference \cite{Richardson2018} a little more detail is provided. The steady-state inhibitory flux can be written
\be
\SJi=-\SRi\int_{v}^\infty dw P(w) e^{-(v-w)/\ami}
\ee
where it should be remembered that $\ami<0$. Well above the unstable fixed point the voltage varies as $\SP(w)\simeq(\Sr\tau/\dT)e^{-(w-\vT)/\dT}$. Substituting in this form and
performing the integral gives an approximation valid at large voltages:
\be
\hspace{-7mm}\SJi&\simeq& \SRi\frac{\ami}{\dT-\ami}\Sr\tau e^{-(v-\vT)/\dT}\simeq \SRi\frac{\ami}{\dT-\ami} \Sr \EscT(v)
\ee
In the second form, the escape time $\EscT(v)$ has been substututed in by noting that $f\sim \dT e^{(v-\vT)/\dT}/\tau$ so that $\EscT\sim \tau e^{-(v-\vT)/\dT}$ at large voltages. Performing the integral transform (see Eq. \ref{TtransformSS}) gives the high-frequency asymptotic
\be
\Mr\simeq\MRi\frac{\Sr\tau}{i\w\tau}\frac{\ami}{\dT-\ami} \label{hricurr}
\ee
for modulation of the inhibitory rate modulation. These results are presented in Fig. \ref{fig3}.

%%%%%%%%%%%%%%%%%%%%%%%%%%%%%%%%%%%%%%%%%%%%%%%%%%%%%%%%%%%%%%%%%%%
%%%%%%%%%%%%%%%%%%%%%%%%%%%%%%%%%%%%%%%%%%%%%%%%%%%%%%%%%%%%%%%%%%%
% Appendix: numerical rate-response for the LIF
%%%%%%%%%%%%%%%%%%%%%%%%%%%%%%%%%%%%%%%%%%%%%%%%%%%%%%%%%%%%%%%%%%%
%%%%%%%%%%%%%%%%%%%%%%%%%%%%%%%%%%%%%%%%%%%%%%%%%%%%%%%%%%%%%%%%%%%
\subsection*{Numerics for the EIF with additive shot noise} \vspace*{-2mm}
In cases where the analytical solution cannot be found in convenient closed form or even for reasons of numerical convenience, the solutions can be found for the steady-state rate or firing-rate modulation using the same method as for the conductance-based case described in Appendix B. The two differences are: first, the forms of the flux equations so that Eqs. (\ref{IJeJi}) are used instead of Eqs. (\ref{GJeJi}); and second, the position of the lower bound $\vlb$. Because there is no inhibitory reversal potential, the lower bound $\vlb$ can be placed at some sufficiently low value such that it does not have a material effect on the results (i.e. in a sufficiently hyperpolarised region where the probability density is very low). Julia code \cite{Bezanson2017} for the EIF with additive shot noise is also provided in the Supplemental Material \cite{MySupp}.

%%%%%%%%%%%%%%%%%%%%%%%%%%%%%%%%%%%%%%%%%%%%%%%%%%%%%%%%%%%%%%%%%%%
%%%%%%%%%%%%%%%%%%%%%%%%%%%%%%%%%%%%%%%%%%%%%%%%%%%%%%%%%%%%%%%%%%%
%%%%%%%%%%%%%%%%%%%%%%%%%%%%%%%%%%%%%%%%%%%%%%%%%%%%%%%%%%%%%%%%%%%
%%%%%%%%%%%%%%%%%%%%%%%%%%%%%%%%%%%%%%%%%%%%%%%%%%%%%%%%%%%%%%%%%%%
% Appendix C
%%%%%%%%%%%%%%%%%%%%%%%%%%%%%%%%%%%%%%%%%%%%%%%%%%%%%%%%%%%%%%%%%%% 
%%%%%%%%%%%%%%%%%%%%%%%%%%%%%%%%%%%%%%%%%%%%%%%%%%%%%%%%%%%%%%%%%%%
%%%%%%%%%%%%%%%%%%%%%%%%%%%%%%%%%%%%%%%%%%%%%%%%%%%%%%%%%%%%%%%%%%%
%%%%%%%%%%%%%%%%%%%%%%%%%%%%%%%%%%%%%%%%%%%%%%%%%%%%%%%%%%%%%%%%%%%
\section*{APPENDIX C: SIMULATIONS} \vspace*{-2mm}
Simulations of the voltage dynamics (Eq. \ref{dvdt}) with synaptic-amplitude distributions drawn from Eqs. (\ref{hexp}) or (\ref{Aexp}) were performed using a forward Euler scheme to provide a comparison to analytical or numerically exact results. All simulations were run at a time step of $0.01$ms and averaged such that results were of the order of the symbol size in figure panels. For the oscillatory responses the amplitudes of the modulated rates for the LIF model were $\MRe\!=\!0.025$kHz and $\MRi\!=\!0.075$kHz for both conductance-based and current-based modulations. For the EIF model these parameters were $0.075$kHz in all cases.

%%%%%%%%%%%%%%%%%%%%%%%%%%%%%%%%%%%%%%%%%%%%%%%%%%%%%%%%%%%%%%%%%%%
%%%%%%%%%%%%%%%%%%%%%%%%%%%%%%%%%%%%%%%%%%%%%%%%%%%%%%%%%%%%%%%%%%%
%%%%%%%%%%%%%%%%%%%%%%%%%%%%%%%%%%%%%%%%%%%%%%%%%%%%%%%%%%%%%%%%%%%
% REFERENCES
%%%%%%%%%%%%%%%%%%%%%%%%%%%%%%%%%%%%%%%%%%%%%%%%%%%%%%%%%%%%%%%%%%% 
%%%%%%%%%%%%%%%%%%%%%%%%%%%%%%%%%%%%%%%%%%%%%%%%%%%%%%%%%%%%%%%%%%%
%%%%%%%%%%%%%%%%%%%%%%%%%%%%%%%%%%%%%%%%%%%%%%%%%%%%%%%%%%%%%%%%%%%

\end{document}